\documentclass[preprint,12pt]{elsarticle}

\usepackage{amssymb}

\usepackage{array}
\usepackage{algorithm2e}

\journal{Future Generation Computer Systems}

\begin{document}

\begin{frontmatter}



\title{A framework for large-scale distributed AI search across disconnected
heterogeneous infrastructures}


\author{Lars Kotthoff, Tom Kelsey and Martin McCaffery}

\address{School of Computer Science\\
University of St Andrews\\
KY16 9SX, United Kingdom}

\begin{abstract}
We present a framework for a large-scale distributed eScience Artificial Intelligence
search. Our approach is generic and can be used for many different problems.
Unlike many other approaches, we do not require dedicated machines, homogeneous
infrastructure or the ability to communicate between nodes. We give special
consideration to the robustness of the framework, minimising the loss of effort
even after total loss of infrastructure, and allowing easy verification of every
step of the distribution process. In contrast to most eScience applications, the
input data and specification of the problem is very small, being easily given in
a paragraph of text. The unique challenges our framework tackles are related to
the combinatorial explosion of the space that contains the possible solutions
and the robustness of long-running computations. Not only is the time
required to finish the computations unknown, but also the resource requirements
may change during the course of the computation. We demonstrate the
applicability of our framework by using it to solve a challenging and hitherto
open problem in computational mathematics. The results demonstrate that our
approach easily scales to computations of a size that would have been impossible
to tackle in practice just a decade ago.

\end{abstract}


\end{frontmatter}


\section{Introduction}
\label{intro}

The last decade has seen an unprecedented rise in the computing power that
institutions and even individuals have access to. This is not only true for
individual processors, but also the number of processors and machines. During
the last few years, a dramatic paradigm shift from ever faster processors to an
ever increasing number of processors and processing elements has occurred. Even
basic contemporary machines have several generic processing elements and
specialised chips for e.g. \ graphics processing.

The size of problems people are interested in solving and the amount of data
that needs to be processed in order to do that has grown dramatically as well.
Today, amounts of data are routinely processed that could not even have been
stored a decade ago. All this presents computer science with new and challenging
research directions.

The processing of so-called ``big data'' is one of the directions where a lot of
research has been done and a lot of tools have been developed. Applications can
be scaled across hundreds of machines relatively easily. The situation in many
areas of Artificial Intelligence is completely different however. Distributing
problems across several machines has been a research endeavour long before the
advent of easily accessible computational resources and big data. The problems
AI aims to solve have always required a large amount of computational resources
to solve problems of practical relevance.

Considering the keen interest of AI researchers in parallelisation, it is
somewhat paradoxical that frameworks to distribute AI techniques are still in
their infancy when it comes to practical applications. One such example is
Apache Mahout~\cite{Mahout}, which leverages the
generic Hadoop framework to distribute Machine Learning algorithms. For AI
search on the other hand, there are, to the best of our knowledge, no similar
frameworks.

Artificial Intelligence search has close links with eScience research, being
used to plan workflows~\cite{AIgrid}, identify optimal protein and DNA
structures~\cite{Mann09,KK10}, and obtain qualitative models of dynamics systems
arising in a wide range of scientific
areas~\cite{Clancy1998,Escrig2002,Radke-Sharpe1998,Menzies1997,KL12}. 

AI search involves the efficient creation, exploration and pruning of very large
search trees (for the game of chess, the tree has an estimated $10^{47}$ nodes).
In many cases it is acceptable to find the first solution from many candidates,
or accept sub-optimal solutions with respect to a cost function to limit the
amount of search performed. However, we often require either all solutions to a
given problem, or a solution that has a guarantee of optimality.

Even when only the first solution is required, the time to find it can quickly
grow to days, months or even years on a single computer. In most cases, this is
unacceptable -- we must be able to find a solution in less time. There are two
strategies for achieving this. The AI search techniques can be improved to be
more efficient for the problem or the search can be distributed across several
machines such that the time to find a solution decreases without actually
decreasing the total effort. The framework presented in this paper pursues the
latter strategy.

Our requirements for such a framework can be summarised as follows.
\begin{itemize}
\item \textbf{Scalability.}\hfill\\
    We want to be able to use as many resources as possible at
    the same time, regardless of type and location and with minimal connectivity
    requirements.
\item \textbf{Robustness.}\hfill\\
    The framework must be able to cope with hardware and similar
    failures. In particular, the amount of computational effort lost because of
    such an event should be small.
\item \textbf{Verifiability.}\hfill\\
    In order to be useful for solving open problems, we must
    be able to follow each step in the distribution process to verify that AI
    search proceeded correctly and no solutions were lost.
\end{itemize}

In this paper we describe a framework that fulfils these requirements. The
design and implementation is motivated by the Recovery Oriented
Computing~\cite{ROC1,ROC2} aspects of the much wider research into Ultralarge
systems~\cite{ULS}. The AI search undertaken is Constraint Programming,
described in Section~\ref{CP}. This is not a restriction, as most AI search
problems can be expressed as Constraint Programming problems. The application
area that we use to evaluate the implementation of the framework is described in
Section~\ref{SG}.

\subsection{Constraint Programming}
\label{CP}

Constraints are a natural and compact way of representing problems that are
ubiquitous in everyday life. Constraint Programming investigates techniques for
solving problems that involve constraints. Common application domains include
other areas of Artificial Intelligence such as planning, but also real world and
industrial applications such as scheduling, design and configuration or
diagnosis and testing. Wallace~\cite{wallace_practical_1996} gives an early
overview of application areas.

Formally, a constraint problem is a triple $(\mathcal{X}, \mathcal{D},
\mathcal{C})$, where $\mathcal{X}$ is a finite indexed set of variables $x_1,
x_2, \ldots, x_n$. Each variable $x_i$ has a finite domain of possible values
$D_i\in\mathcal{D}$. The set $\mathcal{C}$ is a finite set of constraints on the
variables in $\mathcal{X}$. A constraint is a relation that restricts the values
of the variables in its scope. A \emph{solution} to a constraint problem is a
complete assignment of values from the respective domains to all variables
$x_i\in\mathcal{X}$ such that none of the constraints $c_j\in\mathcal{C}$ is
violated. 

In constraint programming, a distinction is usually made between constraint
satisfaction problems (CSPs) and constrained optimisation problems (COPs). A
solution to the former only has to satisfy all the constraints, whereas a
solution to the latter is also given a score by a cost function that needs to be
optimised. As such, it is usually not sufficient to find only the first solution
of a COP even if only one solution is required unless this first solution can be
shown to be optimal. In the remainder of this paper we consider, without loss
of generality, CSPs.

Constraint problems are typically solved by building a search tree in which the
nodes are assignments of values to variables and the edges lead to assignment
choices for the next variable. If at any node a constraint is violated,
search backtracks by returning to a previous state. If a leaf is reached and no
constraints are violated, all variables have been assigned values and this set
of assignments denotes a solution to the CSP.

Clearly the search trees are exponential in the number of variables. Exploring
all of them is infeasible in many cases and inference is used at each node of
the search tree to prune values from the domains of unassigned variables that
cannot be part of a solution based on the assignments made so far. Inference
also allows to backtrack before a constraint is violated -- if the domain of a
particular variable becomes empty, the set of assignments made so far cannot be
part of a solution.

The inference checks have a computational cost and the trade-off is between the
effort of making checks -- hopefully resulting in a reduction of the search
space -- and the effort of searching a presumably larger tree but at a cheaper
cost per node. This is an area of active research and the Handbook of Constraint
Programming~\cite{CSP} provides more details on the many techniques that can be
used to solve constraint problems.

Constraint problems are often highly symmetric. Symmetries may be inherent in
the problem or be created in the process of representing the problem as a CSP.
A symmetry can be as simple as being able to swap the assignments of two
variables in every solution or involve complex permutations of the assignments.
In general, it is desirable to rule out the symmetries during search. This often
leads to a massive reduction in the search space while the solutions that have
been ruled out can be recovered after the problem has been solved at a low
computational cost.

The process of removing symmetries is referred to as symmetry breaking. It
introduces additional constraints that are redundant with respect to the
original problem specification, but rule out symmetrical solutions. More details
on symmetries and symmetry breaking techniques can again be found in the
Constraint Programming Handbook~\cite{CSP}.

\setlength{\extrarowheight}{.3em}
\begin{table}[htb]
\caption{A Semigroup of order $10$.}
\label{table:1}
\begin{center}
\normalsize
\begin{tabular}{|c|cccccccccc|}
\hline
* &  0 &  1 &  2 &  3 &  4 &  5 &  6 &  7 &  8 &  9 \\\hline
0 &  0  &  0  &   0  &   0  &  4  &   4  &  0  &  0  &   4  &   4  \\
1 &  0  &  1  &   0  &   0  &  4  &   4  &  0  &  0  &   4  &   4  \\
2 &  2  &  2  &   2  &   2  &  5  &   5  &  2  &  2  &   5  &   5  \\
3 &  2  &  2  &   2  &   3  &  5  &   5  &  2  &  2  &   5  &   5 \\
4 &  0  &  0  &   0  &   0  &  4  &   4  &  4  &  4  &   0  &   0  \\
5 &  2  &  2  &   2  &   2  &  5  &   5  &  5  &  5  &   2  &   2  \\
6 &  0  &  0  &   2  &   2 &  4  &   5  &  6  &  7  &   8  &   9  \\
7 &  0  &  0  &   2  &   2  &  4  &   5  &  7  &  6  &   9  &   8  \\
8 &  2  &  2  &   0  &   0  &  5  &   4  &  8  &  9  &   7  &   6  \\
9 &  2  &  2  &   0  &   0  &  5  &   4  &  9  &  8  &   6  &   7 \\\hline
\end{tabular}
\end{center}
\end{table}

\subsection{Semigroups}
\label{SG}

We apply our framework to finding the semigroups of order 10. 
A semigroup $T = (S,*)$ consists of a set of elements $S$ and a binary operation
$*: S \times S \rightarrow S$ that is \emph{associative}, satisfying $(x*y)*z =
x*(y*z)$ for each $x,y,z\in S$. 
Table~\ref{table:1} is an illustrative example of such an object.
Given a permutation $\pi$ of the elements of $\{0,\ldots,9\}$, a semigroup
\emph{isomorphic} to $T$ is obtained by permuting the rows, the columns, and
finally the values according to $\pi$. An \emph{anti-isomorphism} is the
transpose of an isomorphism.

The problem addressed in this paper is finding all ways of filling in a blank
table such that multiplication is associative up to symmetric equivalence, i.e.\
up to isomorphism or anti-isomorphism. For orders less than $10$, this problem
can be solved by a combination of enumeration formulae and computation on a
single processor. Table~\ref{table:2} -- with entries taken from sequence
A001423 of the On-Line Encyclopaedia of Integer Sequences -- demonstrates the
combinatoric growth in the number of solutions with increasing order, and
motivates the use of multiple compute nodes to explore the solution space. The
table for the semigroups of order $n$ has $n^2$ cells, and each of these can
take any one of $n$ values. Hence the search space for order $n$ is $n^{n^2}$.
For the problem under consideration, $n = 10$, the size of the search space is
$10^{100}$. To put this number into context, it is currently estimated that
there are approximately $10^{80}$ atoms in the universe. The search space for
our problem is so vast that we cannot possibly hope to solve it by brute force
search.

\begin{table}[htb]
\caption{Number of semigroups of order $n$, considered to be equivalent when they are isomorphic or anti-isomorphic}
\label{table:2}
\begin{center}
\normalsize
\begin{tabular}{|c|r|}
\hline
$n$ & Semigroups \\ \hline
1 & 1\\
2 & 4\\
3 & 18\\
4 & 126\\
5 & 1,160\\
6 & 15,973\\
7 & 836,021\\
8 & 1,843,120,128\\
9 & 52,989,400,714,478\\
 \hline
\end{tabular}
\end{center}
\end{table}

Recent advances in the theory of finite
semigroups have led to an enumerative formula~\cite{DM12} that gives the number
of `almost all' semigroups of given order. Despite this, 256,587,290,511,904
non-equivalent solutions had to be found using the framework described in this
paper.

The constraint model of semigroups of order 10 makes extensive use of the
\emph{element} constraint on natural numbers $N$, $M$ and $P$
$$ N = \langle M_0, \ldots, M_{n-1} \rangle [P]$$
which requires that $N$ is the $P$th element of the list $\langle M_0, \ldots,
M_{n-1} \rangle$ in any solution. This constraint is implemented in many CSP
solvers, including the one developed in our group, Minion~\cite{minion}.

We let $\mathcal{X}_1 = \{ T_{a,b} \mid 0 \leq i,j \leq 9 \}$ be variables
representing the entries in a $10 \times 10$ multiplication table $T$, and
$\mathcal{X}_2
= \{ A_{a,b,c} \mid 0 \leq a,b,c \leq 9 \}$ the variables representing each of
the products of three elements. Our basic CSP contains the variables
$\mathcal{X} = \mathcal{X}_1 \cup \mathcal{X}_2$,
each with domain $D = \{0,\ldots,9\}$. For each triple $(a,b,c)$ of
values from $D$, we post the pair of constraints
$$
\langle T_{a,0},\ldots,T_{a,9}\rangle [T_{b,c}] = A_{a,b,c} = \langle
T_{0,c},\ldots,T_{9,c}\rangle [T_{a,b}]
$$
which enforce associativity. We rule out search for semigroups given by a
formula by posting constraints that require at least one assignment of all the
variables in $\mathcal{X}_2$ to be non-zero. A full description of the CSP model
and its reduction into case-splits is given in~\cite{DJKK12}. 

Finding all solutions of this CSP solves our problem apart from ruling out
symmetric equivalents. Our symmetry group is the set of permutations of
$\{0,\ldots,9\}$ combined with possible transpositions of the tables. If $g =
(\pi,\phi)$ is such a symmetry and $T$ is a multiplication table, then $T^g$ is
the table obtained by first permuting the rows and columns of $T$ according to
$\pi$, and either transposing the table or doing nothing, depending on $\phi$. 

We ensure that only canonical solutions are returned by identifying the symmetry group using the GAP computational algebra package~\cite{GAP4}, then  posting ``lex-leader'' symmetry-breaking
constraints before search. This is a well-known technique for dealing with
symmetries in CSPs~\cite{CGLR96,KLR2004}, made harder to implement in our case because
our symmetries involve both variables and values and made harder to deploy
because we need to post up to $2 \times 10! = 7,257,600$ symmetry-breaking
constraints.

\section{Related work}
\label{relwork}
The parallelisation of depth-first search has been the subject of much
research in the past. The first papers on the subject study the distribution
over various specific hardware architectures and investigate how to achieve good
load balancing~\cite{rao_parallel_1987,kumar_parallel_1987}. Distributed solving
of constraint problems specifically was first explored only a few years
later~\cite{collin_feasibility_1991}.

Backtracking search in a distributed setting has also been investigated by
several authors~\cite{rao_efficiency_1993,sanders_better_1995}. A special
variant for distributed scenarios, asynchronous backtracking, was proposed
in~\cite{yokoo_distributed_1992}. Yokoo et al.\ formalise the distributed
constraint satisfaction problem and present algorithms for solving
it~\cite{yokoo_distributed_1998}.

Schulte presents the architecture of a system that uses
networked computers~\cite{schulte_parallel_2000}. The focus of his approach is
to provide a high-level and reusable design for parallel search and achieve a
good speedup compared to sequential solving rather than good resource
utilisation. More recent papers have explored how to transparently parallelise
search without having to modify existing code~\cite{michel_parallelizing_2007}.

Most of the existing work is concerned with the problem of effectively
distributing the workload such that every compute node is kept busy. The most
prevalent technique used to achieve this is work stealing. The compute nodes
communicate with each other and nodes which are idle request a part of the work
that a busy node is doing. Blumofe and Leiserson propose and discuss a work
stealing scheduler for multithreaded computations
in~\cite{blumofe_scheduling_1999}. Rolf and Kuchcinski investigate different
algorithms for load balancing and work stealing in the specific context of
distributed constraint solving~\cite{rolf_load-balancing_2008}.

Several frameworks for distributed constraint solving have been proposed and
implemented, e.g.\ FRODO~\cite{frodo}, DisChoco~\cite{dischoco} and
Disolver~\cite{disolver}. All of these approaches have in common that the
systems to solve constraint problems are modified or augmented to support
distribution of parts of the problem across and communication between multiple
compute nodes. The constraint model of the problem remains unchanged however; no
special constructs have to be used to take advantage of distributed solving. All
parallelisation is handled in the respective solver. This does not preclude the
use of an entirely different model of the problem to be solved for the
distributed case in order to improve efficiency, but in general these solvers
are able to solve the same model both with a single executor and distributed
across several executors.

The decomposition of constraint problems into subproblems which can be solved
independently has been proposed in~\cite{michel_decomposition-based_2004},
albeit in a different context. In this work, we explore the use of this
technique for parallelisation. A similar approach was taken
in~\cite{rolf_load-balancing_2008}, but requires parallelisation support in the
solver.

\section{Distributing CSPs}
\label{distcsp}

Our approach to parallelising the solving of constraint problems has been
previously described in~\cite{kotthoff_distributed_2010}. This paper updates the
description and, crucially, reports results from an application of the
framework.

Constraint problems are typically solved by searching through the possible
assignments of values to variables. After each such assignment, inference can
rule out possible future assignments based on past assignments and
the constraints. This process builds a search tree that explores the space of
possible (partial) solutions to the constraint problem.

There are two different ways to build up these search trees -- $n$-way branching
and $2$-way branching. This refers to the number of new branches which
are explored after each node. In $n$-way branching, all the $n$ possible
assignments to the next variable are branched on. In $2$-way branching, there
are two branches. The left branch is of the form $x=y$ where $x$ is a variable
and $y$ is a value from its domain. The right branch is of the form $x\neq y$.

The more commonly used way is $2$-way branching, implemented for example in
the Minion constraint
solver~\cite{minion}, available at \verb+http://minion.sf.net+. However,
regardless of the way the branching is done, exploring the branches can be done
concurrently. No information between the branches needs to be exchanged in order
to find a solution to the problem.

We exploit this fact by, given the model of a constraint problem, generating new
models which partition the remaining search space. These models can then be
solved independently. We furthermore represent the state of the search by adding
additional constraints such that the splitting of the model can occur at any
point during search. The new models can be resumed, taking advantage of both the
splitting of the search space and the search already performed.

\subsection{Model splitting}

Our new approach to the distributed solving of constraint problems requires the
constraint solver to modify the constraint model but does not require explicit
parallelisation support in the solver.

To split the remaining search space of a constraint problem, we signal the
solver to stop. Now we partition the domain for the variable currently under
consideration into $n$ pieces of roughly equal size. Then we create $n$ new
models and to each in turn add constraints ruling out the other $n-1$ partitions
of that domain. Each one of these models restricts the possible assignments to
the current variable to one $n$th of its domain.

As an example, consider the case $n=2$. The variable under consideration is $x$
and its domain is $\{1,2,3,4\}$. We generate $2$ new models. One of them has the
constraint $x\leq 2$ added and the other one $x\geq 3$. Thus, solving the first
model will try the values $1$ and $2$ for $x$, whereas the second model will try
$3$ and $4$.

The main problem when splitting constraint problems into parts that can be
solved in parallel is that the size of the remaining search space for each of
the splits is impossible to predict reliably. This directly affects the
effectiveness of the splitting however -- if the search space is distributed
unevenly, some of the workers will be idle while the others do most of the work.

Our approach allows to repeatedly split the search space after search has
started. We use the procedure described above several times, each time adding
more constraints to the model. In addition, we add \emph{restart nogoods}, that
is, additional constraints that tell the solver how much of the search space has
been explored. Constraints added in a previous iteration are not affected by
constraints added later -- regardless of how often we split, no parts of the
search space will be ``lost'', potentially missing solutions. Similarly, no part
of the search space will be visited repeatedly.

\begin{figure*}[htb]
\begin{center}
\includegraphics[width=.66\textwidth]{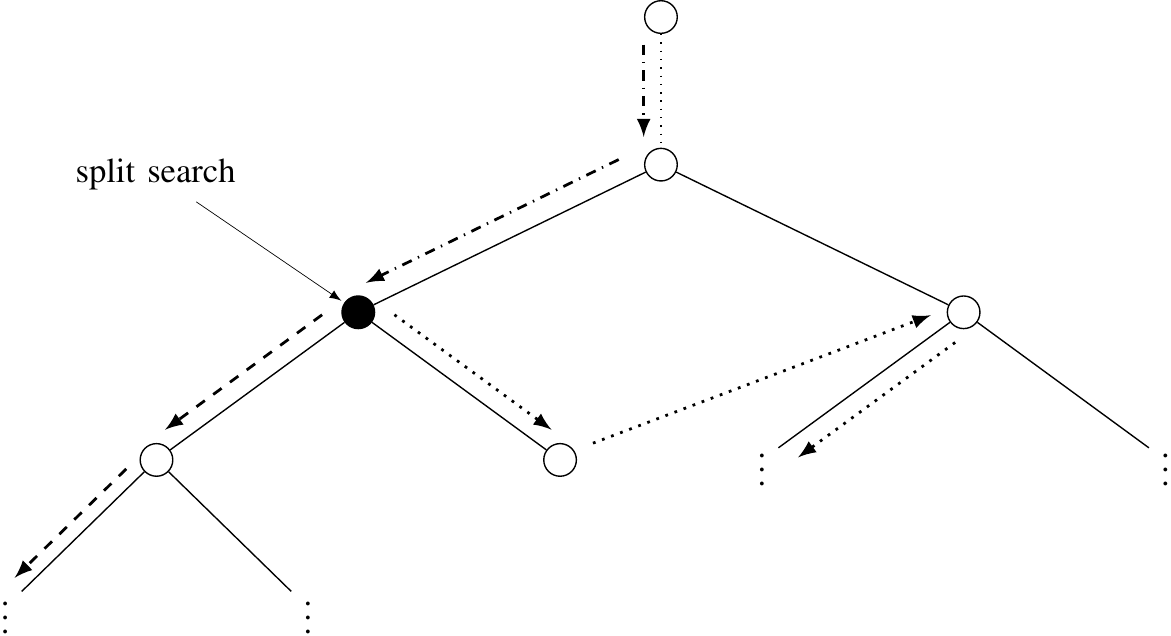}
\end{center}
\caption{Illustration of how search proceeds with splitting. The dashed and
dotted line shows the search up to the solid black node where the models are
split. The nodes the two parallel searches explored subsequently are shown with
dashed and dotted lines, respectively.}
\label{fig:split}
\end{figure*}

Assume for example that we are doing $2$-way branching, the variable currently
under consideration is again $x$ with domain $\{1,2,3,4\}$ and the branches that
we have taken to get to the point where we are are $x\neq 1$ and $x\neq 2$. The
generated new models will all have the constraints $x\neq 1$ and $x\neq 2$ to
get to the point in the search tree where we split the problem. Then we add
constraints to partition the search space based on the remaining values in the
domain of $x$ similar to the previous example. The splitting process and
subsequent parallel search is illustrated in Figure~\ref{fig:split}.

Using this technique, we can create new chunks of work whenever a worker becomes
idle by simply asking one of the busy workers to split the search space. The
search is then resumed from where it was stopped and the remaining search space
is explored in parallel by the two workers. Note that there is a runtime
overhead involved with stopping and resuming search because the constraints
which enable resumption must be taken into account and the solver needs to
explore a small number of search nodes to get to the point where it was stopped
before. There is also a memory overhead because the additional constraints need
to be stored.

We have implemented this approach in a development version of Minion, which we
are planning to release to the public. Experiments show that the
overhead of stopping, splitting and resuming is not significant for
large problems.

In practice, we run Minion for a specified amount of time, then stop, split and
resume instead of splitting at the beginning and when workers become idle. This
approach is much simpler and works well for large problems. The algorithm is
detailed in Procedure~\ref{distalgo}. It creates an $n$-ary split tree of
models for $n$ new models generated at each split. The procedure for finding all
solutions is similar. Initially, the potential for distribution is small but
grows exponentially as more and more search is performed. We have found that
$n=2$ works well in practice because it is the easiest to implement and
minimises the number of models created.

\setlength{\algomargin}{0pt}
\SetAlFnt{\normalsize}
\begin{procedure}
\SetKwFunction{solved}{solved?}
\SetKwFunction{exhausted}{search space exhausted?}
\SetKwInOut{Input}{Input}
\SetKwInOut{Output}{Output}
\SetKwFor{For}{for}{do in parallel}{end}

\Input{constraint problem $X$, allotted time $T_{max}$ and splitting factor
$n\geq 2$}
\Output{a solution to $X$ or ``no solution'' if no solution has been found}
\BlankLine
run the constraint solver with input $X$ until termination or $T_{max}$\;
\BlankLine
\uIf{\solved{$X$}}{
   terminate workers\;
   \Return{solution}\;
}
\uElseIf{\exhausted}{
    \Return{``no solution''}\;
}
\Else{
   $X' \leftarrow X$ with new constraints ruling out search already
   performed\;
   split $X'$ into $n$ parts $X'_1, \ldots, X'_n$\;
   \BlankLine
   \For{$i \leftarrow 1$ \KwTo $n$}{
       distSolve($X'_n$, $T_{max}$,$n$)\;
   }
}
\caption{distSolve($X$,$T_{max}$,$n$): Recursive procedure to find the first
solution to a constraint problem distributed across several workers.}
\label{distalgo}
\end{procedure}

Minion models are stored in ordinary files. Each time the search space is split,
two new input files are written. We modified the output produced by Minion to
include the names of the files it produced and included the name of the file
that was run when the search space was split in the new model files. This way,
we can easily trace the splitting of the search space across the different
files.

\subsection{Comparison to existing approaches}

The main advantages of our approach are as follows.
\begin{itemize}
\item We require only minimal modifications to existing constraint solvers. In
    particular, we do not require network communication and work stealing to be
    implemented.
\item We do not require communication between workers to achieve good
    utilisation.
\item The creation of separate model files when splitting increases the
    robustness against worker failure and provides accountability for every
    step.
\end{itemize}

For the purposes of a framework for solving large Artificial Intelligence search
problems, the last point is especially crucial. The nature of the applications
that we have in mind is such that it will be neither easy to verify whether a
solution is valid nor feasible to repeat the calculations to get a confirmation.
Furthermore, we have to be able to rely on the capability to recover from
failures without having to repeat all the work.

By creating regular ``snapshots'' of the search done, the resilience against
failure increases. This is in contrast to most other approaches, where the
reliability of the system is decreased by using techniques that distribute work
and rely on several machines instead of just a single one. Such systems have
then to take additional measures to mitigate the problems caused by failures of
machines or communication links. Every time we split the search space, the
modified models are saved. As they contain constraints that rule out the search
already done, we only lose the work done after that point if a worker fails.
This means that the maximum amount of work we lose in case of a total failure of
all workers is the allotted time $T_{max}$ times the number of workers $|w|$.

We note that our approach provides many of the advantages of efforts dedicated
to improving the robustness and accountability of computations,
e.g.~\cite{benabdelkader_provenance_2011}, but is much easier to implement and
only requires a minimal amount of supporting infrastructure.

Another consequence of our approach is that the solving process can be moved to
a different set of workers after it has been started without losing any work.
This may become necessary if parts of the problem require much more memory to
solve than other parts. Instead of provisioning workers with a large number of
resources for the entire duration of the computation, it becomes feasible to do
this on-demand. This allows for excellent and easy integration with existing
services that offer on-demand computing, such as a cloud.

\subsection{Large-scale distribution}

In the previous sections, we have described the techniques that enable the
distribution of the solving of a constraint problem across a set of workers, but
not the system to take care of the actual distribution. The implementation of
such a system is notoriously difficult, hence we decided to leverage a
tried-and-tested existing system.

For the purposes of a framework that allows to distribute problems across a
large number of heterogeneous workers, the Condor HPC
system~\cite{thain_distributed_2005} is particularly suitable. It runs in many
different operating and network environments and provides most of the
functionality we require out of the box. In particular, it allows for the
transfer of files that are created on the worker back to the master -- the
constraint models that split the search space.

Condor allows work units to be submitted to a central node which puts them in a
queue to be executed when a worker becomes available. In our case, a constraint
model is a unit of work and splitting the search space on one of the workers
creates two new units of work that are transferred back to the master and queued
for execution. The condor job submission system makes sure that a job is
executed to completion, i.e.\ if a worker node fails while it is processing a
work unit, Condor requeues the work unit and sends it to a different worker.

Each Condor work unit needs to be created separately. In order to submit models
that split the search space and are created during search, we have implemented a
custom control system that monitors Condor and takes the appropriate action when
split models are returned. The control system is an almost trivial piece of
software that was very easy to implement -- all of the heavy lifting is done by
Condor.

\bigskip

While Condor is a very adequate system for our needs, its installation is not
always straightforward. Ultimately, the scale of problems we are aiming for
might require not thousands of machines but tens of thousands. No institution or
even set of institutions has sufficient resources to make this available for a
single project. Fortunately, the rise of the internet has facilitated so-called
volunteer computing, where interested users can ``donate'' compute time to a
project of their choice.

The best-known framework for such projects is BOINC, the Berkeley Open
Infrastructure for Network Computing~\cite{anderson_boinc_2004}. It has been
used for many applications, including astrophysics, biology and mathematics. We
have integrated the Minion constraint solver with the BOINC framework in a way
that allows for splitting the search space. This system provides many of the
benefits of Condor but makes it much easier for non-technical users to contribute.

\begin{figure*}[htb]
\includegraphics[height=\textwidth,angle=-90]{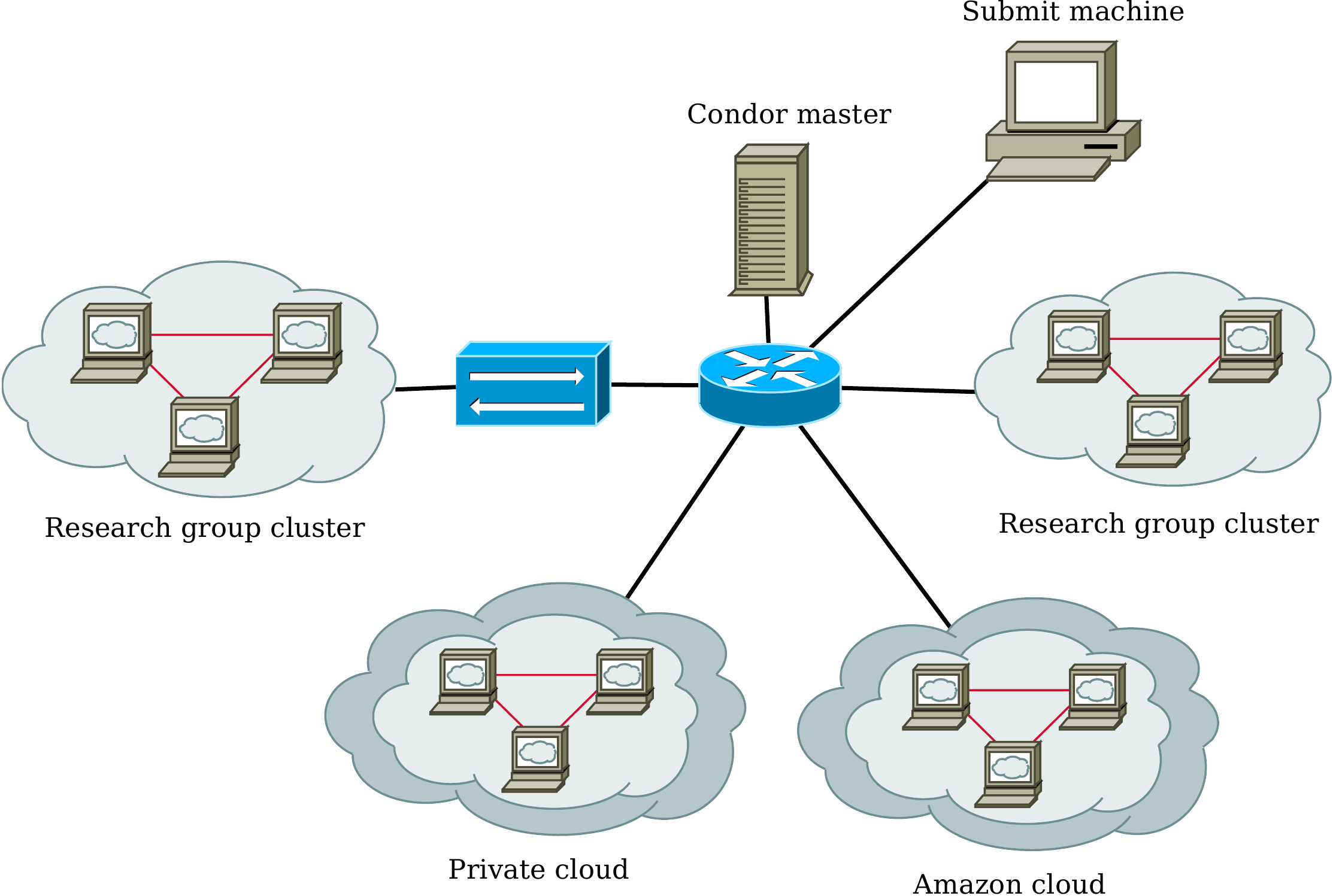}
\caption{Overview of resources used for the enumeration of semigroups. We used
two research clusters: a private cloud hosted in St Andrews and the Amazon
cloud. One of the research clusters was behind a NAT switch such that no
machines on the outside could reach it directly and all connections had to be
initiated from within it.}
\label{fig:nodes}
\end{figure*}

\section{Application and discussion}
\label{app}

We first validated our framework empirically by using it to compute the number
of semigroups of order~9, a problem that had previously been solved using
non-distributed search. We were able to confirm the known result on a number of
different hardware configurations and splitting parameters, i.e.\ the time
search is run before splitting the model.

Encouraged by the results of these experiments, we started the calculation of
the number of semigroups of order~10. The hardware configuration throughout the
computations varied, but the principal resources we used are shown in
Figure~\ref{fig:nodes}. Here, one of the main advantages of our framework became
apparent. The different resources we used were located in different networks
that did not always have unrestricted connectivity to the other nodes. One of
the research group clusters for example was behind a NAT in its own private
network and unable to receive connections from outside this network. We were
still able to utilise the resources to their full extent.

The submit machine and the Condor master shown in Figure~\ref{fig:nodes} were
not used for any of the computations, but only for the management of the
calculations. It should be noted that there is no reason to have dedicated
machines for those purposes as the resource requirements for the tasks they
performed were very low. In principle, a machine used for management of the
computations could also be used to perform computations itself.

The maximum number of processors that we used in parallel at any one time was
about~150. One of the reasons for using the Amazon cloud was that it turned out
that the machines we had available locally did not have enough memory to explore
some parts of the search space efficiently. We were able to move those
calculations to virtual machines in the Amazon cloud with suitable
specifications and seamlessly integrate the results of those computations with
the rest.

The total CPU time we expended to solve the problem (i.e.\ find exactly
256,587,290,511,904 semigroups from~$10^{100}$ potential tables) was
approximately~133 years. This effort was achieved in approximately~18 months;
full details of the mathematics and the case-splits used are described
in~\cite{DJKK12}. The limiting factor were the resources that were available to
us. Even though we did not start with a short search time before splitting,
enough split models to utilise all our resources were available after a few
hours. For shorter computations, it might be desirable to facilitate faster
splitting at the beginning to achieve good utilisation earlier, but for our
purposes the framework as described previously was sufficient. The number of
split models produced suggested that we could have utilised up to several
thousand processors to a very high degree.

The robustness of our framework proved useful several times during the
computations. Events that we successfully coped with included power and network
outages, air-conditioning failures, physical machines being switched off and
virtual machines disappearing. The damage in terms of computational effort lost
was very limited in all cases. Condor was able to recover from most of these
failures without any manual intervention by simply re-queueing the failed jobs.
The verification of the distribution process revealed that because of the
re-queueing a small part of the search space had been explored several times,
but we were able to isolate and discard the duplicate model and output files.

After the computations finished, we were able to verify each step of the
distribution and solving process. Therefore, we are confident that the result we
obtained is correct. Ultimately, certainty of the correctness can only be
established by either a new mathematical model that allows to calculate the
computed number directly, or by independent verification through a second
computation.

\section{Conclusions and future work}
\label{concs}

We have presented a framework for the large-scale distribution of AI search in
constraint programming across resources with minimal network connectivity
requirements. We have implemented this framework and applied the implementation
to solving a hitherto open problem in computational mathematics. Throughout this
application, the framework has proved to fulfill all our requirements. It is
capable of scaling almost seamlessly to a large number of distributed and
heterogeneous resources while minimising losses due to hardware failures. It
furthermore provides the functionality to verify each step of the distribution
process, creating confidence in the results.

The type of our application is relatively rare in eScience. Instead of large
amounts of data to process, we have a very concise problem specification that
takes vast computational resources to solve. We believe that the nature of such
problems presents unique challenges to eScience that have rarely been considered
so far.

There is no indication that the positive experiences we have had with the
specific application described here is limited to that
particular problem. We have, neither in the design of the framework nor its
application, made any assumptions to that effect. We are currently evaluating
the application of the framework to other problems that can be expressed as
constraint problems and require large computational efforts.

An obvious avenue for future work apart from the application to new problems
that we would like to explore is the evaluation of the implementation of the
framework that uses BOINC instead of Condor. An application to the same problem
would allow us to not only judge the differences in terms of distribution
effectivity and utilisation, but also to independently verify the results that
we have obtained. While we are confident that we would indeed obtain the same
result, an empirical verification would eliminate any doubts about this aspect
of the framework.

We are planning to release as open source the modifications we have made to the
Minion constraint solver in order to support splitting searches. Furthermore, we
are intending to release all other components of the framework that are not
already available to the public,  thus enabling other researchers to tackle
similarly large problems and providing a framework that we hope will prove
useful to the research community.

\section*{Acknowledgments}

The authors thank Chris Jefferson for useful discussions on the implementation
of the framework. Tom Kelsey is supported by UK EPSRC grant EP/H004092/1. Lars
Kotthoff is supported by an EPSRC fellowship.

Parts of the computational resources for this project were provided by an Amazon
Web Services research grant. We thank the School of Computer Science, the Centre
for Interdisciplinary Research in Computational Algebra and Cloud Co-laboratory
(all of the University of St Andrews) for providing additional computational
resources.





\bibliographystyle{model1-num-names}
\bibliography{KKM}







\end{document}